\PassOptionsToPackage{unicode}{hyperref}
\PassOptionsToPackage{hyphens}{url}
\PassOptionsToPackage{dvipsnames,svgnames,x11names}{xcolor}
\documentclass[
  a4paper,
  headings=standardclasses,
  chapterprefix=false,
  parskip=half]{scrartcl}
\usepackage{xcolor}
\usepackage{amsmath,amssymb}
\usepackage{iftex}
\ifPDFTeX
  \usepackage[T1]{fontenc}
  \usepackage[utf8]{inputenc}
  \usepackage{textcomp} % provide euro and other symbols
\else % if luatex or xetex
  \usepackage{unicode-math} % this also loads fontspec
  \defaultfontfeatures{Scale=MatchLowercase}
  \defaultfontfeatures[\rmfamily]{Ligatures=TeX,Scale=1}
\fi
\usepackage{lmodern}
\ifPDFTeX\else
\fi
\IfFileExists{upquote.sty}{\usepackage{upquote}}{}
\IfFileExists{microtype.sty}{% use microtype if available
  \usepackage[]{microtype}
  \UseMicrotypeSet[protrusion]{basicmath} % disable protrusion for tt fonts
}{}
\makeatletter
\@ifundefined{KOMAClassName}{% if non-KOMA class
  \IfFileExists{parskip.sty}{%
    \usepackage{parskip}
  }{% else
    \setlength{\parindent}{0pt}
    \setlength{\parskip}{6pt plus 2pt minus 1pt}}
}{% if KOMA class
  \KOMAoptions{parskip=half}}
\makeatother
\makeatletter
\ifx\paragraph\undefined\else
  \let\oldparagraph\paragraph
  \renewcommand{\paragraph}{
    \@ifstar
      \xxxParagraphStar
      \xxxParagraphNoStar
  }
  \newcommand{\xxxParagraphStar}[1]{\oldparagraph*{#1}\mbox{}}
  \newcommand{\xxxParagraphNoStar}[1]{\oldparagraph{#1}\mbox{}}
\fi
\ifx\subparagraph\undefined\else
  \let\oldsubparagraph\subparagraph
  \renewcommand{\subparagraph}{
    \@ifstar
      \xxxSubParagraphStar
      \xxxSubParagraphNoStar
  }
  \newcommand{\xxxSubParagraphStar}[1]{\oldsubparagraph*{#1}\mbox{}}
  \newcommand{\xxxSubParagraphNoStar}[1]{\oldsubparagraph{#1}\mbox{}}
\fi
\makeatother

\usepackage{longtable,booktabs,array}
\usepackage{multirow}
\usepackage{calc} % for calculating minipage widths
\usepackage{etoolbox}
\makeatletter
\patchcmd\longtable{\par}{\if@noskipsec\mbox{}\fi\par}{}{}
\makeatother
\IfFileExists{footnotehyper.sty}{\usepackage{footnotehyper}}{\usepackage{footnote}}
\makesavenoteenv{longtable}
\usepackage{graphicx}
\makeatletter
\newsavebox\pandoc@box
\newcommand*\pandocbounded[1]{% scales image to fit in text height/width
  \sbox\pandoc@box{#1}%
  \Gscale@div\@tempa{\textheight}{\dimexpr\ht\pandoc@box+\dp\pandoc@box\relax}%
  \Gscale@div\@tempb{\linewidth}{\wd\pandoc@box}%
  \ifdim\@tempb\p@<\@tempa\p@\let\@tempa\@tempb\fi% select the smaller of both
  \ifdim\@tempa\p@<\p@\scalebox{\@tempa}{\usebox\pandoc@box}%
  \else\usebox{\pandoc@box}%
  \fi%
}
\def\fps@figure{htbp}
\makeatother

\NewDocumentCommand\citeproctext{}{}
\NewDocumentCommand\citeproc{mm}{%
  \begingroup\def\citeproctext{#2}\cite{#1}\endgroup}
\makeatletter
 \let\@cite@ofmt\@firstofone
 \def\@biblabel#1{}
 \def\@cite#1#2{{#1\if@tempswa , #2\fi}}
\makeatother
\newlength{\cslhangindent}
\newlength{\csllabelwidth}
\newenvironment{CSLReferences}[2] % #1 hanging-indent, #2 entry-spacing
 {\begin{list}{}{%
  \setlength{\itemindent}{0pt}
  \setlength{\leftmargin}{0pt}
  \setlength{\parsep}{0pt}
  \ifodd #1
   \setlength{\leftmargin}{\cslhangindent}
   \setlength{\itemindent}{-1\cslhangindent}
  \fi
  \setlength{\itemsep}{#2\baselineskip}}}
 {\end{list}}
\usepackage{calc}

\usepackage{booktabs}
\usepackage{adjustbox}
\usepackage{dcolumn}
\usepackage{hyperref}
\usepackage{orcidlink}
\usepackage{authblk}
\newcommand{\sym}[1]{\ifmmode^{#1}\else\(^{#1}\)\fi}
\newcommand{\emaillink}[1]{\href{mailto:#1}{\nolinkurl{#1}}}

\makeatletter
\@ifpackageloaded{float}{}{\usepackage{float}}
\floatstyle{plain}
\@ifundefined{c@chapter}{\newfloat{apptbl}{h}{loapptbl}}{\newfloat{apptbl}{h}{loapptbl}[chapter]}
\floatname{apptbl}{Table A}
\newcommand*\quartoapptblref[1]{Table \hyperref[#1]{A\ref{#1}}}
\@ifpackageloaded{caption}{}{\usepackage{caption}}
\DeclareCaptionLabelFormat{quartoapptblreflabelformat}{#1#2}
\makeatother
\makeatletter
\@ifpackageloaded{float}{}{\usepackage{float}}
\floatstyle{plain}
\@ifundefined{c@chapter}{\newfloat{appfig}{h}{loappfig}}{\newfloat{appfig}{h}{loappfig}[chapter]}
\floatname{appfig}{Figure A}
\newcommand*\quartoappfigref[1]{Figure \hyperref[#1]{A\ref{#1}}}
\@ifpackageloaded{caption}{}{\usepackage{caption}}
\DeclareCaptionLabelFormat{quartoappfigreflabelformat}{#1#2}
\makeatother
\makeatletter
\@ifpackageloaded{caption}{}{\usepackage{caption}}
\AtBeginDocument{%
\ifdefined\contentsname
  \renewcommand*\contentsname{Table of contents}
\else
  \newcommand\contentsname{Table of contents}
\fi
\ifdefined\listfigurename
  \renewcommand*\listfigurename{List of Figures}
\else
  \newcommand\listfigurename{List of Figures}
\fi
\ifdefined\listtablename
  \renewcommand*\listtablename{List of Tables}
\else
  \newcommand\listtablename{List of Tables}
\fi
\ifdefined\figurename
  \renewcommand*\figurename{Figure}
\else
  \newcommand\figurename{Figure}
\fi
\ifdefined\tablename
  \renewcommand*\tablename{Table}
\else
  \newcommand\tablename{Table}
\fi
}
\@ifpackageloaded{float}{}{\usepackage{float}}
\floatstyle{ruled}
\@ifundefined{c@chapter}{\newfloat{codelisting}{h}{lop}}{\newfloat{codelisting}{h}{lop}[chapter]}
\floatname{codelisting}{Listing}

\makeatother
\makeatletter
\usepackage{pdflscape}
\makeatother
\makeatletter
\@ifpackageloaded{tikz}{}{\usepackage{tikz}}
\makeatother
\makeatletter
\@ifpackageloaded{caption}{}{\usepackage{caption}}
\@ifpackageloaded{subcaption}{}{\usepackage{subcaption}}
\makeatother
\usepackage{bookmark}
\IfFileExists{xurl.sty}{\usepackage{xurl}}{} % add URL line breaks if available
\hypersetup{
  pdftitle={Collaboration for the Bioeconomy},
  pdfauthor={Philipp Jonas Kreutzer; Josef Taalbi},
  colorlinks=true,
  linkcolor={blue},
  filecolor={Maroon},
  citecolor={Blue},
  urlcolor={Blue},
  pdfcreator={LaTeX via pandoc}}

\title{Collaboration for the Bioeconomy}
\subtitle{Evidence From Innovation Output in Sweden, 1970--2021}
\author{%
Philipp Jonas Kreutzer%
\textsuperscript{%
1%
}%
\thanks{Corresponding author: philipp\_jonas.kreutzer@ekh.lu.se}%
\,\orcidlink{0000-0003-4234-9043}%
, %
Josef Taalbi%
\textsuperscript{%
1%
}%
\,\orcidlink{0000-0002-2560-9504}%
}
\affil[1]{%
Department of Economic History, %
Lund University%
, Lund%
, Sweden%
}
\date{2026-02-04}
\begin{document}
\maketitle
\begin{abstract}
Collaboration is expected to play a central role in the transition to a
bioeconomy---a central pillar of a green economy. Such collaboration is
supposed to connect traditional biomass processing firms with diverse
actors in fields where biomass ought to substitute existing or create
novel products and processes. This study analyzes the network of
technology collaborations among innovating firms in Sweden between 1970
and 2021. The results reveal generally positive associations between
direct and indirect ties, with meaningful increases in innovation output
for each additional direct collaboration partner. Relationships between
brokerage positions and innovation output were statistically
insignificant, and cognitive proximity---while following theoretical
expectations---materially insignificant. These associations are mostly
equal between actors heavily invested in the bioeconomy and those
focusing on other innovation areas, indicating that these actors operate
under largely similar mechanisms linking collaboration and subsequent
innovation output. These results suggest that stimulating collaboration
broadly---rather than attempting to optimize collaboration
compositions---could result in higher number of significant Swedish
innovations, for bioeconomy and other sectors alike.

\end{abstract}

\section{Introduction}\label{introduction}

The bioeconomy is expected to play a major role in the transition to
low-carbon economies. Substantial innovation efforts are required to
transition from fossil-derived fuel and material resources to bio-based
substitutes and alternatives
(\citeproc{ref-issa2019BioeconomyExpertsPerspectives}{Issa et al.,
2019}). Mainstream bioeconomy visions argue that these innovations
require integrating traditional biomass sectors (such as forestry) into
diverse downstream industries, where bio-based alternatives must
displace fossil-based products and processes
(\citeproc{ref-holmgren2020BioeconomyImaginariesReview}{Holmgren et al.,
2020}). Yet, evidence suggests that this transition is not happening
fast enough, with structural weaknesses of bioeconomy innovation systems
identified as key bottlenecks
(\citeproc{ref-giurca2017ForestbasedBioeconomyGermany}{Giurca and Späth,
2017}; \citeproc{ref-hellsmark2016InnovationSystemStrengths}{Hellsmark
et al., 2016}).

Policymakers, industry experts and researchers have identified
collaborations between firms of different sizes and collaborations
across industry boundaries as essential to accelerating the bioeconomy
transition
(\citeproc{ref-el-chichakli2016PolicyFiveCornerstones}{El-Chichakli et
al., 2016}). Calls for increased collaboration are compelling:
traditional biomass actors possess raw-material expertise, but are
focused on maintaining competitiveness in traditional, low-margin
product categories
(\citeproc{ref-lamberg2017InstitutionalPathDependence}{Lamberg et al.,
2017}); downstream actors, on the other hand, understand market needs
but lack biomass processing capabilities. Bridging these knowledge gaps
through collaboration should unlock innovation opportunities and
strengthen the innovation system, ultimately resulting in increased
innovation (\citeproc{ref-hekkert2007FunctionsInnovationSystems}{Hekkert
et al., 2007}). However, empirical evidence on whether and how
collaboration actually drives bioeconomy innovation remains limited,
hampering evidence-based policymaking.

Three empirical challenges have hampered evaluating the role of
collaboration in bioeconomy innovation. First, the bioeconomy is
inherently a cross-sectoral and emerging phenomenon, which complicates
delineating actors and activities
(\citeproc{ref-wydra2020MeasuringInnovationBioeconomy}{Wydra, 2020}).
Previous empirical research has sidestepped this issue by focusing on
specific elements, such as biorefineries (for example,
\citeproc{ref-bauer2018InnovationBioeconomyDynamics}{Bauer et al.,
2018}); patent-intensive activities, such as biotechnology (for example,
\citeproc{ref-abbasiharofteh2021StillShadowWall}{Abbasiharofteh and
Broekel, 2021}; \citeproc{ref-cooke2001NewEconomyInnovation}{Cooke,
2001}); or specific sectors, such as pulp and paper
(\citeproc{ref-soderholm2012FirmCollaborationEnvironmental}{Söderholm
and Bergquist, 2012}). But these studies leave the broader bioeconomy
innovation system aside. Second, most collaboration research relies on
patents (for example,
\citeproc{ref-abbasiharofteh2021StillShadowWall}{Abbasiharofteh and
Broekel, 2021}) or R\&D-applications (for example,
\citeproc{ref-hernandez2024ImprovedCausalTest}{Hernandez et al., 2024}),
or combinations thereof
(\citeproc{ref-fornahl2011WhatDrivesPatent}{Fornahl et al., 2011}),
which do not directly capture the crucial innovation step. Third,
distinguishing if bioeconomy actors face fundamentally different
collaboration dynamics requires embedding them in broader innovation
system contexts. This holistic perspective is absent from literature on
bioeconomy subsectors.

This paper investigates the role of collaboration for innovation
activity in the entire Swedish forest-based bioeconomy over five decades
(1970--2021). We construct novel panel data linking commercialized
innovation to collaboration networks and test whether
bioeconomy-intensive firms exhibit different relationships between key
network properties and subsequent innovation than other firms. Data for
innovation output and collaboration stem from a longitudinal innovation
output database, which captures commercialized innovation rather than
proxies (\citeproc{ref-kander2019InnovationTrendsIndustrial}{Kander et
al., 2019}; \citeproc{ref-sjoo2014DatabaseSwedishInnovations}{Sjöö et
al., 2014};
\citeproc{ref-taalbi2025InnovationPatentsInformationtheoretic}{Taalbi,
2025}). This approach addresses three gaps: it covers the entire
forest-based bioeconomy rather than subsectors; it measures innovation
directly, rather than relying on patents or R\&D proxies; and embeds
bioeconomy actors within the broader Swedish innovation system over
multiple decades rather than a few years.

Our findings challenge prevailing assumptions about bioeconomy
collaboration bottlenecks. Although bioeconomy actors produce overall
fewer innovations, we find no evidence that they operate under different
collaboration mechanisms than other actors. Increasing the number of
collaborators a firm has access to consistently increased its predicted
yearly innovation output. However, this appears driven by collaboration
in general; indirect ties and brokerage positions showed no clear
effects. Following theoretical expectations of an inverted-U
relationship, cognitive proximity had negligible practical relevance.
Contrary to popular concerns, bioeconomy firms were not hampered by
being too cognitively similar, falling below the predicted optimum.

These results suggest that low bioeconomy innovation output cannot be
attributed to insufficient collaboration. For policy, this implies that
broad stimuli to increase collaboration may prove more effective than
attempting to create ideal bioeconomy partnerships.

\section{Previous Literature}\label{sec-lit}

This study investigates how collaboration networks shape innovation
outputs in the forest-based bioeconomy. Previous literature raises two
relevant themes. First, how do collaboration networks affect innovation
outcomes in general? And second, do bioeconomy innovations require
different collaboration patterns?

\subsection{Collaboration Networks and
Innovation}\label{collaboration-networks-and-innovation}

Since the seminal work by Burt
(\citeproc{ref-burt1992StructuralHolesSocial}{1992}) and Ahuja
(\citeproc{ref-ahuja2000CollaborationNetworksStructural}{2000}), a
sizable literature has investigated the relationship between
organizations' collaboration networks and innovation outcomes. We
identify three main mechanisms through which networks shape innovation.

First, the number of direct ties between an organization and
collaborating partners is traditionally assumed to have a positive
association with innovation outcomes. Collaborations spur innovation
through sharing of knowledge, complementarity of skills
(\citeproc{ref-arora1990ComplementarityExternalLinkages}{Arora and
Gambardella, 1990}) and knowledge spillovers
(\citeproc{ref-jaffe1986TechnologicalOpportunitySpillovers}{Jaffe,
1986}), and by enabling scale economies in development projects
(\citeproc{ref-ahuja2000CollaborationNetworksStructural}{Ahuja, 2000}).
In addition, since collaborating partners are themselves embedded in
collaboration networks, an organization may benefit indirectly from the
ties of its collaboration partners
(\citeproc{ref-ahuja2000CollaborationNetworksStructural}{Ahuja, 2000}).
In line with these notions, empirical studies have generally found a
positive association between direct and indirect linkages and innovation
outcomes (\citeproc{ref-ahuja2000CollaborationNetworksStructural}{Ahuja,
2000}; \citeproc{ref-wang2014KnowledgeNetworksCollaboration}{Wang et
al., 2014}).

Second, actors may benefit from connecting otherwise disconnected
network subgroups, rather than being constrained to partners who
themselves share connections with one another
(\citeproc{ref-burt2004StructuralHolesGood}{Burt, 2004}). Spanning such
structural holes, between nodes which have little or no mutual
connections should increase access to more (diverse) information and
thus increase subsequent innovation. However, empirical evidence about
the effect of structural holes on innovations is mixed: some work has
demonstrated a positive relationship
(\citeproc{ref-chiu2012StructuralEmbeddednessInnovation}{Chiu and Lee,
2012}; \citeproc{ref-reagans2001NetworksDiversityProductivity}{Reagans
and Zuckerman, 2001}), while others found that firms whose collaboration
networks consisted of nodes with few mutual ties were less innovative
(\citeproc{ref-ahuja2000CollaborationNetworksStructural}{Ahuja, 2000}).

Next, beyond network structure, the \emph{type} of connections matter.
Here, a core concept is absorptive capacity, or the likelihood of
organizations making productive use of shared knowledge within a network
(\citeproc{ref-cohen1990AbsorptiveCapacityNew}{Cohen and Levinthal,
1990}). We focus on one specific factor of absorptive capacity:
cognitive proximity. Cognitive proximity refers to the extent to which
collaboration partners share similar knowledge bases and expertise,
which affects both the likelihood of collaborating and its successful
innovation outcome
(\citeproc{ref-boschma2005ProximityInnovationCritical}{Boschma, 2005}).
While similar knowledge aids communication and decreases the risk of
collaboration and may thus increase productivity, to high cognitive
proximity may make new knowledge recombinations less likely, stifling
innovation
(\citeproc{ref-cantner2007TechnologicalProximityChoice}{Cantner and
Meder, 2007};
\citeproc{ref-nooteboom2007OptimalCognitiveDistance}{Nooteboom et al.,
2007}).

In other words: innovation output is expected to follow an inverted
U-shaped relationship with cognitive proximity
(\citeproc{ref-boschma2005ProximityInnovationCritical}{Boschma, 2005};
\citeproc{ref-broekel2012KnowledgeNetworksDutch}{Broekel and Boschma,
2012}; \citeproc{ref-nooteboom2007OptimalCognitiveDistance}{Nooteboom et
al., 2007}). More formally, Nooteboom et al.
(\citeproc{ref-nooteboom2007OptimalCognitiveDistance}{2007}) suggested
that the absorptive capacity of firms increases with cognitive proximity
between a focal firm and its collaboration partners according to
\(A = a_1 + a_2 CP\), while novelty value decreases with cognitive
proximity \(N = b_1 - b_2 CP\). Thus, the innovation performance of a
firm follows the quadratic equation:

\begin{equation}\phantomsection\label{eq-cp}{I = AN = a_1b_1 -  (a_1 b_2 - a_2 b_1 ) CP  - a_2 b_2 CP^2, }\end{equation}

Empirical support for this relationship is strong, but not unanimous. It
has been identified in joint venture formation involving American firms
(\citeproc{ref-mowery1998TechnologicalOverlapInterfirm}{Mowery et al.,
1998}), explaining patent activity
(\citeproc{ref-fornahl2011WhatDrivesPatent}{Fornahl et al., 2011};
\citeproc{ref-nooteboom2007OptimalCognitiveDistance}{Nooteboom et al.,
2007}), and may complement or even substitute other forms of proximity
such as geography
(\citeproc{ref-garciamartinez2024GeographicalCognitiveProximity}{Garcia
Martinez et al., 2024}). Other studies, such as Heringa et al.
(\citeproc{ref-heringa2014HowDimensionsProximity}{2014}) found overall
positive effects of cognitive proximity on innovation outcomes, but they
did not find evidence of the hypothesized inverse-U pattern, nor did
Broekel and Boschma
(\citeproc{ref-broekel2012KnowledgeNetworksDutch}{2012}) find evidence
for this quadratic relationship.

From these three mechanisms, we formulate the following testable
hypotheses:

\begin{itemize}

    \item [\emph{H1a}] Organizations' direct ties have a positive impact on subsequent innovations.

    \item [\emph{H1b}] Organizations' indirect ties have a positive impact on subsequent innovations.

    \item [\emph{H1c}] Structural holes have a positive impact on subsequent innovations.

    \item [\emph{H1d}] Cognitive proximity has an inverted U-shaped relationship to innovation output (\ref{eq-cp}).
\end{itemize}

\subsection{Collaboration Networks in the
Bioeconomy}\label{collaboration-networks-in-the-bioeconomy}

Generally speaking, collaboration is expected to create, ``\ldots{}
better outcomes and efficiency of innovation \ldots{}'' in the
bioeconomy (\citeproc{ref-issa2019BioeconomyExpertsPerspectives}{Issa et
al., 2019, p. 8}). The argument is straightforward: bioeconomy supply
chains are inherently complex, spanning traditional biomass sectors (for
example, pulp and paper) and downstream sectors to create novel products
(for example, petrochemistry for biofuel refineries)
(\citeproc{ref-guerrero2018CrosssectorCollaborationForest}{Guerrero and
Hansen, 2018};
\citeproc{ref-guerrero2021CompanylevelCrosssectorCollaborations}{Guerrero
and Hansen, 2021}). Gendron
(\citeproc{ref-gendron2024CollaborationInnovationPolicy}{2024}) captured
the prevailing view of bioeconomy experts: ``by fostering collaboration
and maintaining a long-term vision, the bioeconomy can fulfill its
potential to become an integral part of a sustainable future---not just
an alternative but the foundation of the global economy'' (p.~250).

However, empirical work on bioeconomy collaboration reveals several
tensions. On one hand studies in Finland
(\citeproc{ref-laakkonen2023ImplicationsSustainabilityTransition}{Laakkonen
et al., 2023}) and Germany
(\citeproc{ref-bogner2019KnowledgeNetworksGerman}{Bogner, 2019};
\citeproc{ref-stober2023BioeconomyInnovationNetworks}{Stöber et al.,
2023}) found that large, established actors have sought out
collaborations to advance biotechnology and novel applications of
biomass. On the other hand, these actors avoided directly collaborating
with other large actors
(\citeproc{ref-laakkonen2023ImplicationsSustainabilityTransition}{Laakkonen
et al., 2023}), and may even use their central network positions to
advance their traditional, rather than truly transformative agendas
(\citeproc{ref-bogner2019KnowledgeNetworksGerman}{Bogner, 2019};
\citeproc{ref-stober2023BioeconomyInnovationNetworks}{Stöber et al.,
2023}).

Collaboration has played an important historical role in the Swedish
bioeconomy---especially in advancing clean technologies in the pulp and
paper sector
(\citeproc{ref-bergquist2011GreenInnovationSystems}{Bergquist and
Söderholm, 2011};
\citeproc{ref-soderholm2012FirmCollaborationEnvironmental}{Söderholm and
Bergquist, 2012}). Of particular note is the active role of the Swedish
government in establishing, supporting, and even participating in
collaborative platforms. For example through research institutes or
organizations, such as the Sulphate Pulp Committee, founded in
1908-1909, the Swedish Pulp and Paper Research Institute (founded in
1945) and the Swedish Environmental Research Institute (IVL) founded in
1966. These actors have served to promote and concert industrial
activity around important environmental problem complexes
(\citeproc{ref-soderholm2012FirmCollaborationEnvironmental}{Söderholm
and Bergquist, 2012}). But, more recent studies have found insufficient
cross-sectoral collaboration in Sweden and argued that this lack is
among the main reasons for insufficient resource mobilization in the
bioeconomy (\citeproc{ref-bauer2018InnovationBioeconomyDynamics}{Bauer
et al., 2018};
\citeproc{ref-hansen2017UnpackingResourceMobilisation}{Hansen and
Coenen, 2017};
\citeproc{ref-hellsmark2016InnovationSystemStrengths}{Hellsmark et al.,
2016}). For biorefineries, a central pillar of the bioeconomy, Bauer et
al. (\citeproc{ref-bauer2018InnovationBioeconomyDynamics}{2018}) showed
that the Swedish collaboration network favored intra-industry
collaboration, rather than collaboration across sectors. The general
lack of cross-sectoral actor networks may even be a central
transformation challenge
(\citeproc{ref-mossberg2021ChallengesSustainableIndustrial}{Mossberg et
al., 2021}).

A potential explanation for this lack of collaboration can be found in a
preference for high cognitive proximity. Despite long-standing calls for
collaboration as a key to competitive success in Swedish pulp and paper
(\citeproc{ref-melander2005IndustrywideBeliefStructures}{Melander,
2005}), especially to obtain knowledge from petrochemical actors
(\citeproc{ref-ericsson2018ClimateInnovationsPaper}{Ericsson and
Nilsson, 2018}), Hansen
(\citeproc{ref-hansen2010RoleInnovationForest}{2010}) argued that
incumbent actors showed a lack of interest in emerging biorefinery
technologies, focusing rather on existing technologies. German
biotechnology collaboration was similarly associated with higher
cognitive proximity between partners
(\citeproc{ref-roesler2017RoleUniversitiesNetwork}{Roesler and Broekel,
2017}), indicating that this might be a widespread feature of bioeconomy
innovators.

From this literature, one would expect that technology collaboration
within the bioeconomy may be characterized by higher cognitive proximity
than other sectors. In addition, one may expect that collaboration in
the bioeconomy is characterized by too high cognitive proximity,
relative to the optimum level posited in hypothesis \emph{H1d}. We
therefore formulate the bioeconomy specific hypotheses that:

\begin{itemize}

\item [\emph{H2a}] Collaborations within the bioeconomy have higher cognitive proximity than other sectors

and 

\item [\emph{H2b}] Bioeconomy actor's level of cognitive proximity is too high relative to the optimum level.

\end{itemize}

\section{Data and Methods}\label{data-and-methods}

We based our analysis on a novel panel dataset of innovation and
collaboration outcomes, which we constructed from a literature-based
innovation output database covering Swedish innovations between 1970 and
2021. The panel combines innovation output counts with firm and network
characteristics for each actor involved in commercializing at least one
innovation in the study period. We then estimate Poisson panel
regressions, and conduct an instrumental variable estimation as a
sensitivity check for potential endogeneity between firm and network
characteristics and innovation output.

\subsection{Collaborations in Swedish Literature-Based Innovation Output
Data}\label{sec-swinno-data}

Previous studies of the nexus of collaboration and innovation have most
often used proxies of innovation such as patents or R\&D. Similarly, the
bioeconomy has mostly been measured through input factors such as public
or private R\&D spending, enrolled university students or throughput
measures such as patents
(\citeproc{ref-wydra2020MeasuringInnovationBioeconomy}{Wydra, 2020}).
While useful, these measures present well-known limitations, which are
of particular importance for bioeconomy research.

R\&D expenditure does not necessarily capture innovation, but more
importantly for the bioeconomy, it is biased towards research intensive
sectors and firms and thus only accounts for specific,
research-intensive bioeconomy transition pathways
(\citeproc{ref-vivien2019HijackingBioeconomy}{Vivien et al., 2019}).
Similar limitations apply to patents: the propensity to patent differs
across sectors which complicates capturing some types of bioeconomy
innovations related to wood and pulp and paper products
(\citeproc{ref-arundel1998WhatPercentageInnovations}{Arundel and Kabla,
1998}; \citeproc{ref-cohen1990AbsorptiveCapacityNew}{Cohen and
Levinthal, 1990};
\citeproc{ref-taalbi2025InnovationPatentsInformationtheoretic}{Taalbi,
2025}). Moreover, the bioeconomy includes supplier-dominated,
science-based as well as production-intensive industries, all with
different patenting propensities
(\citeproc{ref-pavitt1984SectoralPatternsTechnical}{Pavitt, 1984}).
Ultimately, the perhaps biggest drawback of both R\&D expenditure and
patents is that they capture proxies of innovation rather than
innovation directly. A measure that remedies this shortcoming is the
literature-based innovation output (LBIO) method
(\citeproc{ref-kleinknecht1993LiteraturebasedInnovationOutput}{Kleinknecht
and Reijnen, 1993};
\citeproc{ref-vanderpanne2007IssuesMeasuringInnovation}{van der Panne,
2007}).

In this study, we use a database of Swedish innovation output (SWINNO)
based on trade journals
(\citeproc{ref-kander2019InnovationTrendsIndustrial}{Kander et al.,
2019}; \citeproc{ref-sjoo2014DatabaseSwedishInnovations}{Sjöö et al.,
2014}). The underlying LBIO methodology relies on the selection of
independent trade journals with an explicit aim to inform specialized as
well as the interested general audience about innovation and
technological developments. To ensure that the data is not biased
towards firms with large advertising resources, the methodology only
considers the edited sections of the trade journals, excluding product
announcement and advertisement sections. Hence, only innovations deemed
to have sufficient novelty are reported
(\citeproc{ref-kander2019InnovationTrendsIndustrial}{Kander et al.,
2019}; \citeproc{ref-sjoo2014DatabaseSwedishInnovations}{Sjöö et al.,
2014}).

The journal articles are collected, read and innovations entered into
the database if they fulfill additional criteria. Specifically,
innovations are operationalized as new products and processes that are
brought to market. In addition, innovations must be developed by at
least one Swedish organization.

\begin{figure}

\centering{

\input{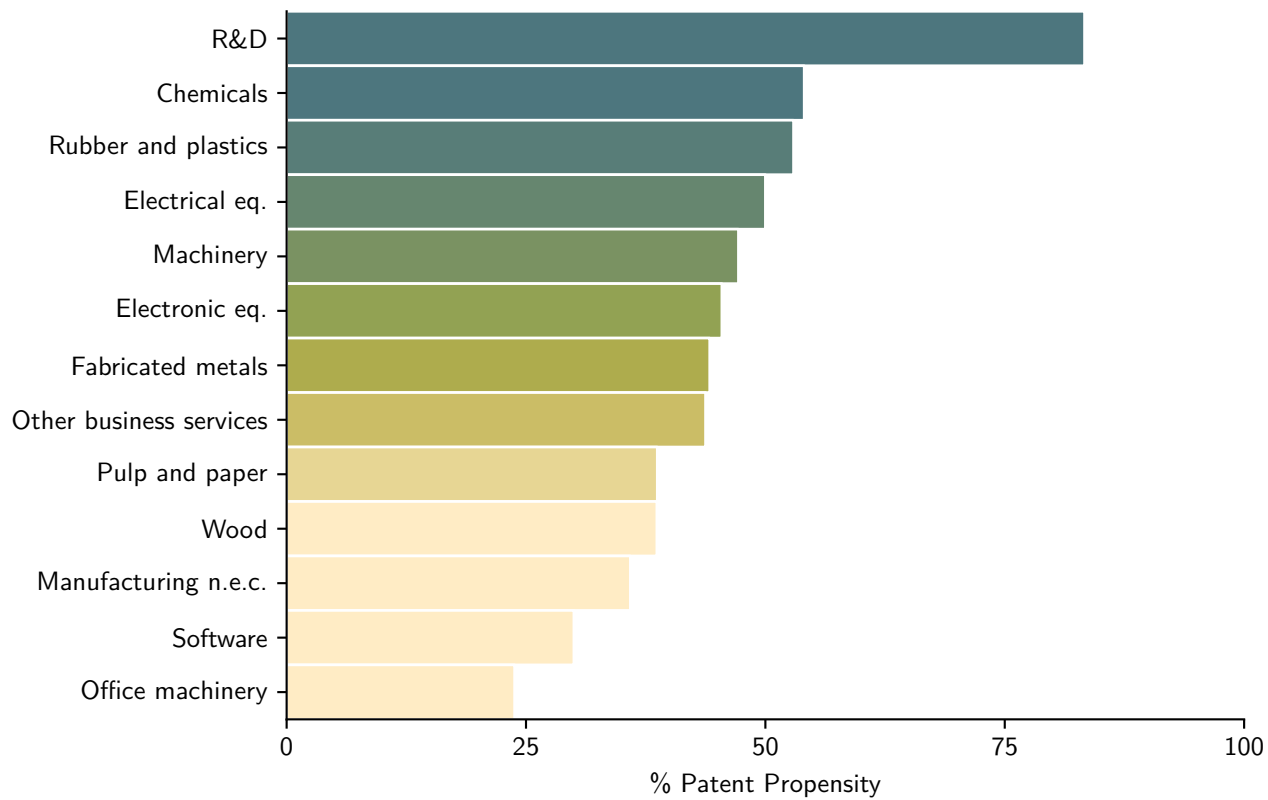}

}

\caption{\label{fig-patent-propensity}\textbf{Patent Propensity of
Bioeconomy Innovations by Sector.} Calculations based on Johansson et
al. (\citeproc{ref-johansson2022LinkingInnovationsPatents}{2022}) and
Taalbi
(\citeproc{ref-taalbi2025InnovationPatentsInformationtheoretic}{2025}).}

\end{figure}

Between 1970 and 2021, we identified 4,972 innovation in the database.
More than half of the innovations captured were never patented
(\citeproc{ref-taalbi2025InnovationPatentsInformationtheoretic}{Taalbi,
2025}). Within the bioeconomy, we see similar patterns.
Figure~\ref{fig-patent-propensity} shows the share of bioeconomy
innovations that were patented across sectors with at least 5
innovations. In a few sectors, such as R\&D innovations and chemicals, a
majority of innovations were patented. However, in key sectors of the
forest-based bioeconomy only about 38\% innovations were patented, and
below 30\% were patented in computers and software.

\subsubsection{Collaboration Data}\label{collaboration-data}

Trade journal articles not only mention actors who commercialized an
innovation, but also entities involved in their development. This
enables construction of collaboration networks based on joint innovation
production. The collaboration network constructed in this work hence
reflects technology collaborations in which firms and organizations
actively developed an innovation, but excludes other types of commercial
networks that firms may have access to, such as joint ventures or lobby
organizations.

Our data relies on explicit information in trade journals articles. The
vast majority of articles have an ``innovation focus'', providing
in-depth insights into the history of the innovation and report not only
the name of the firm(s), but also the persons involved in the
innovation's development. Shorter articles may occasionally focus on the
commercializing firm and the product itself, rather than its
development, which could potentially lead to underreporting
collaborators. However, these types of articles are rare as the LBIO
methodology requires the article to describe the innovation in enough
detail to infer novelty and commercialization status
(\citeproc{ref-sjoo2014DatabaseSwedishInnovations}{Sjöö et al., 2014}).

There are two main limitations of the data used in this study. First,
our innovation indicator measures significant innovations and does not
capture incremental product developments or product variations.
Secondly, our data captures only those technology collaborations that
led to at least one commercialized innovation. Collaborations that fail
to yield a commercialized innovation remain unobserved. While such
efforts may also produce experience and knowledge, the focus on
successful outcomes ensures that the exchange we observe is significant
for the firms.

\subsection{Bioeconomy Operationalization}\label{sec-bioeconomy}

Empirical measures of the bioeconomy suffer from definitional ambiguity
and the cross-sectoral scope
(\citeproc{ref-jander2020MonitoringBioeconomyTransitions}{Jander et al.,
2020}; \citeproc{ref-wydra2020MeasuringInnovationBioeconomy}{Wydra,
2020}). Our approach to identify innovating bioeconomy firms departs
from first identifying bioeconomy innovations, and then classifying
firms based on bioeconomy innovation intensity.

\subsubsection{Bioeconomy Innovation}\label{bioeconomy-innovation}

We identified bioeconomy innovations through a combination of
origination or usage in key forest-biomass sectors (forestry, wood
products, pulp and paper manufacturing) and keyword matching
forest-bioeconomy value chains in innovation descriptions
(\citeproc{ref-wolfslehner2016ForestBioeconomyNew}{Wolfslehner et al.,
2016}). The list of sectors and keywords is reproduced in the Appendix
(Table~\ref{tbl-query}). We manually cleaned the results to remove false
positives and thus identified forest-based bioeconomy 653 (13.13\% of
the 4,972 innovations in included in this study).

\subsubsection{Bioeconomy Organizations}\label{bioeconomy-organizations}

Since bioeconomy innovation transcends individual sectors, statistical
taxonomies of economic activity alone cannot identify bioeconomy
organizations. Instead, we identified firms based on past activity: an
organization participating in the bioeconomy is an organization whose
past cumulative innovation activities included at least 25\% bioeconomy
innovation. This threshold captures organizations with a substantial,
but not necessarily exclusive, engagement in the bioeconomy, and
accounts dynamically for the emerging and cross-sectoral bioeconomy
phenomenon. As robustness checks, we used more restrictive thresholds,
and report results for 50\% of innovations in the Appendix
(Table~\ref{tbl-panel-threshold-robustness}).

\subsection{Network Construction and Variable
Definition}\label{sec-network-construction}

We constructed undirected collaboration networks, \(G(N, L)\), based on
all innovations commercialized between 1970 and 2021. A collaboration
network \(G\) consists of nodes \(N\), representing named entities and
links \(L\), representing collaborations between distinct nodes. In the
network, a node refers to a unique named organization involved in at
least one innovation. Each link contains information about the year of
the collaboration and a unique identifier tying the collaboration to a
specific innovation. Table~\ref{tbl-network-descriptives} displays key
characteristics of the total network, the subcomponent of bioeconomy
innovations and the network excluding bioeconomy innovations.

\begin{table}

\caption{\label{tbl-network-descriptives}Descriptive Statistics of
Sweden's Innovation Collaboration Network}

\centering{

\begin{tabular*}{\linewidth}{@{\extracolsep{\fill}}lrrr}
\toprule
 & Total Network & Only Bioeconomy & Excluding Bioeconomy \\ 
\midrule\addlinespace[2.5pt]
Edges & 2,202 & 294 & 1,152 \\
Nodes & 1,479 & 244 & 1,101 \\
Average Degree & 2.98 & 2.41 & 2.09 \\
SD Degree & 6.16 & 2.08 & 2.13 \\
Min Degree & 1 & 1 & 1 \\
Max Degree & 111 & 11 & 28 \\
\bottomrule
\end{tabular*}
\begin{minipage}{\linewidth}
Note: Network excludes isolates. Only bioeconomy includes only bioeconomy innovations. Excluding bioeconomy removes nodes with at least one bioeconomy innovation.\\
\end{minipage}

}

\end{table}%

A crucial issue when working with organizational data is the level of
aggregation. In SWINNO, entities involved in developing the innovation
are recorded as they are mentioned in the source. For independent
organizations this poses no issues. However, for organizations that are
part of a larger corporation, locating the innovation comes with a
potential measurement error. While it is possible to aggregate
subsidiary organizations into parent organizations, not all articles
mention specific subsidiaries. In these cases the innovation is recorded
as originating from the parent company. A related issue emerges from the
temporal span of this research. Between 1970 and 2021 some companies,
universities and other organizations have changed owners, names, or
both. Following the logic of Hylmö and Taalbi
(\citeproc{ref-hylmo2024RiseFallGiants}{2024}) and Taalbi
(\citeproc{ref-taalbi2026LongrunPatternsDiscovery}{2026}) we
consolidated organizations accounting for name varieties, well-known
name changes and aggregation into major corporate groups.

\subsection{Network Position
Variables}\label{network-position-variables}

Table~\ref{tbl-variables} summarizes all variables used in this study.
Our main variables, testing hypothesis \emph{H1a}--\emph{H1d} are direct
ties (testing \emph{H1a}), indirect ties (\emph{H1b}), two-step
betweenness (\emph{H1c}) and cognitive proximity (\emph{H1d}). To test
whether collaborations within the bioeconomy have different patterns of
cognitive proximity (hypotheses \emph{H2a} and \emph{H2b}), we interact
cognitive proximity with whether a firm belongs to the bioeconomy.

\begin{table}

\caption{\label{tbl-variables}Summary of Operationalization of Concepts
and Hypotheses}

\centering{

\begin{tabular}{p{5cm}p{7cm}p{3cm}}
\hline
Name & Description & Hypothesis \\
\hline
\emph{Dependent variable} &  & \\
Innovation count & Count of innovations of firm $i$ in year $t$ & \\
\emph{Independent variables} &  &\\
Direct ties & Number of collaborators of focal firm $i$ in year $t-1$ & 
\emph{H1a} \\
Indirect ties & Number of collaborators of focal firm $i$ in year $t-1$ & \emph{H1b} \\
% at distance $d$
Two-Step Betweenness & Betweenness centrality restricted to geodesics of length two, in year $t-1$ & \emph{H1c}  \\
Cognitive proximity & Normalized Jaccard index of the economic activity of a focal firm and its collaborators, in year $t-1$ & \emph{H1d, H2a, H2b}  \\
Bioeconomy & Firms having made at least 25 \% of their innovations in the bioeconomy, year $t-1$  & \\
\emph{Control variables} &  & \\
Age & Years since firm was founded at year $t-1$ & \\
Year & Set of dummy variables associated with the year under analysis &\\
\hline
\end{tabular}

}

\end{table}%

\subsubsection{Direct and Indirect Ties}\label{sec-ties}

To measure direct ties, we used the degree centrality of a node. For
indirect ties, defined as collaboration partners of a direct
collaboration partner, we used a simple count of neighbors of node \(i\)
at distance two as \(f_i^2\). For example, a node with one direct
collaboration partner who in turn has two collaboration partners other
than the source node has one direct tie and two indirect ties.

\subsubsection{Brokerage}\label{brokerage}

Structural holes measure the extent to which a node bridges otherwise
unconnected parts of a network, thus filling in a ``hole'' in the
network. Organizations that fill these holes are in a unique position to
broker knowledge between otherwise disconnected communities
(\citeproc{ref-burt2004StructuralHolesGood}{Burt, 2004}); a task
expected to be especially important for the bioeconomy. Following
Everett and Borgatti
(\citeproc{ref-everett2020UnpackingBurtsConstraint}{2020}) we employed
two-step betweenness rather than Burt
(\citeproc{ref-burt2004StructuralHolesGood}{2004})'s constraint. Burt's
constraint suffers from two limitations in our context: it is undefined
for isolates (requiring their exclusion); and, including both size and
constraint measures in regression introduces multicollinearity and
removes the validity of Burt's constraint as a measure for structural
holes (\citeproc{ref-everett2020UnpackingBurtsConstraint}{Everett and
Borgatti, 2020}). Two-step betweenness calculates the extent to which a
node serves as exclusive intermediary between nodes at distance two
(\citeproc{ref-borgatti2006GraphtheoreticPerspectiveCentrality}{Borgatti
and Everett, 2006};
\citeproc{ref-brandes2008VariantsShortestpathBetweenness}{Brandes,
2008}).

\subsubsection{Cognitive Proximity}\label{sec-knowledge-measure}

To test our hypotheses regarding cognitive proximity, we used a
Jaccard-index measure of the overlap between organizations' competences.
Our measure of knowledge bases rests on two-digit Swedish Standard
Classification (SNI) codes reported by the organizations to Swedish
registries, as well as two-digit SNI codes classifying each previous
innovation output. We homogenized all SNI codes to 2002 equivalents.

We gathered self-reported register data of organizations economic
activity from Statistics Sweden and, for the period 1997 to 2021 by
matching nodes to register data published in Serrano, a database of
company level financial history
(\citeproc{ref-weidenmanperSerranoDatabaseAnalysis}{Weidenman, Per,
n.d.}). This data reflects classifications at the entity level for all
years in which the entity reported its activity classification. To
address discrepancies between parent and subsidiary reportings, we
aggregate SNI codes to the highest organizational level under our
aggregation scheme. A parent company thus inherits SNI codes from all of
its subsidiaries rather than solely receiving administrative
classification.

Additionally, each SWINNO innovation is given SNI product code. This
allows us to measure knowledge across time based on previous innovation
output. We aggregated this data as well to the highest organizational
level. Consider a company that produces a manufacturing innovation in a
year \(t\) and one chemical innovation in year \(t+1\). In year \(t\) no
information about previous innovation expertise is available. In \(t+1\)
we know that the company has expertise in manufacturing. In \(t+2\) we
know that the company has broadened it to include chemistry in addition
to machinery.

We integrate these measures to leverage their complementary strengths.
Register data provides initial knowledge proxies before an actor's first
innovation, while innovation-based measures track dynamic knowledge
development. Operationally, register knowledge for year \(t\) is
attributed to year \(t-1\), with subsequent updates incorporating
innovation-based knowledge accumulation. To illustrate, consider an
organization registered with the Swedish statistical office as a pulp
and paper company. For its first innovation we infer that it drew on
knowledge stemming from pulp and paper. Assuming again that the first
innovation is a new machine, we update the companies knowledge base to
include both pulp and paper as well as machinery.

Using our measure of knowledge, we calculated the cognitive proximity
between collaboration partners based Jaccard-index measure according to:

\begin{equation}\phantomsection\label{eq-jaccard}{ J(A, B) = \frac{|A \cap B|}{|A \cup B|},}\end{equation}

where, \(A\) and \(B\) represent the knowledge bases of node \(i\) and
its neighbor \(j\), respectively
(\citeproc{ref-boschma2005ProximityInnovationCritical}{Boschma, 2005})
By definition, the Jaccard-index would be 0 if one of the collaboration
partners' knowledge sets were empty. Imputing knowledge from register
data mitigates this, ensuring that firms innovating for the first time
do not skew the measure by definition.

\subsubsection{Controlling for Firm Age}\label{controlling-for-firm-age}

To control for additional firm characteristics, we focused on the age of
the organizations. Age is highly correlated with organizational size,
resource availability and cumulative innovation output. We constructed
age from two data sources to account for our aggregation of firms. For
major organizations, entry and exit years were manually collected and
reconciled with our aggregation scheme. Further, we could extract
registration dates and operational years from the Serrano database
(\citeproc{ref-weidenmanperSerranoDatabaseAnalysis}{Weidenman, Per,
n.d.}). For unmatched organizations, we imputed entry and exit with
ten-year windows around first and last observed innovations, consistent
with median differences between registration and first innovation (8
years) and last innovation and exit (9 years) in the matched sample.

\subsection{Econometric Specification}\label{econometric-specification}

\subsubsection{Panel Construction}\label{panel-construction}

We constructed an unbalanced panel with annual observations for active
firms. Network measures were calculated from slices of the network \(g\)
up to the year before we observed innovation activity:
\(g_{[1970, t-1]}\). For example, for innovations commercialized in the
year 1975, we calculated network statistics for the node based on a
network from 1970 to 1974. Table~\ref{tbl-panel-desc} shows the means
and standard deviations for all variables used in the main regressions
for the whole sample as well as for bioeconomy and other firms. We
present robustness checks in the Appendix, further restricting the
sample to include only firms which collaborated at least once
(Table~\ref{tbl-firm-robustness}) and increasing the network window to
20 years (Table~\ref{tbl-panel-network-robustness}).

\begin{landscape}

\begin{table}

\caption{\label{tbl-panel-desc}Descriptive Statistics of Panel Data}

\centering{

\begin{tabular*}{\linewidth}{@{\extracolsep{\fill}}lrrrrrrrrr}
\toprule
 & \multicolumn{3}{c}{Full Panel} & \multicolumn{3}{c}{Bioeconomy Firms Only} & \multicolumn{3}{c}{Excluding Bioeconomy Firms} \\ 
\cmidrule(lr){2-4} \cmidrule(lr){5-7} \cmidrule(lr){8-10}
 & Count & Mean & SD & Count & Mean & SD & Count & Mean & SD \\ 
\midrule\addlinespace[2.5pt]
Innovation count & 89,931 & 0.065 & 0.299 & 7,816 & 0.042 & 0.246 & 82,115 & 0.067 & 0.303 \\
Direct Ties & 89,931 & 0.362 & 1.393 & 7,816 & 0.573 & 1.355 & 82,115 & 0.342 & 1.395 \\
Indirect Ties & 89,931 & 0.966 & 4.600 & 7,816 & 1.237 & 4.669 & 82,115 & 0.940 & 4.593 \\
2-step Betweenness & 89,931 & 8.270 $\times$ 10\textsuperscript{-6} & 1.596 $\times$ 10\textsuperscript{-4} & 7,816 & 3.703 $\times$ 10\textsuperscript{-6} & 3.290 $\times$ 10\textsuperscript{-5} & 82,115 & 8.705 $\times$ 10\textsuperscript{-6} & 1.667 $\times$ 10\textsuperscript{-4} \\
Cognitive proximity & 89,931 & 0.051 & 0.157 & 7,816 & 0.093 & 0.207 & 82,115 & 0.047 & 0.151 \\
Age & 89,931 & 25.206 & 39.358 & 7,816 & 35.440 & 40.720 & 82,115 & 24.232 & 39.087 \\
Bioeconomy Firm & 89,931 & 0.087 & 0.282 & 7,816 & 1.000 & 0.000 & 82,115 & 0.000 & 0.000 \\
\bottomrule
\end{tabular*}
\begin{minipage}{\linewidth}
Note: All variables---except for innovation count---are once-lagged. Sample restricted to firm-year observations with available lagged predictors. Subgroups defined by bioeconomy status at t-1.\\
\end{minipage}

}

\end{table}%

\end{landscape}

\begin{figure}

\centering{

\pandocbounded{\includegraphics[keepaspectratio]{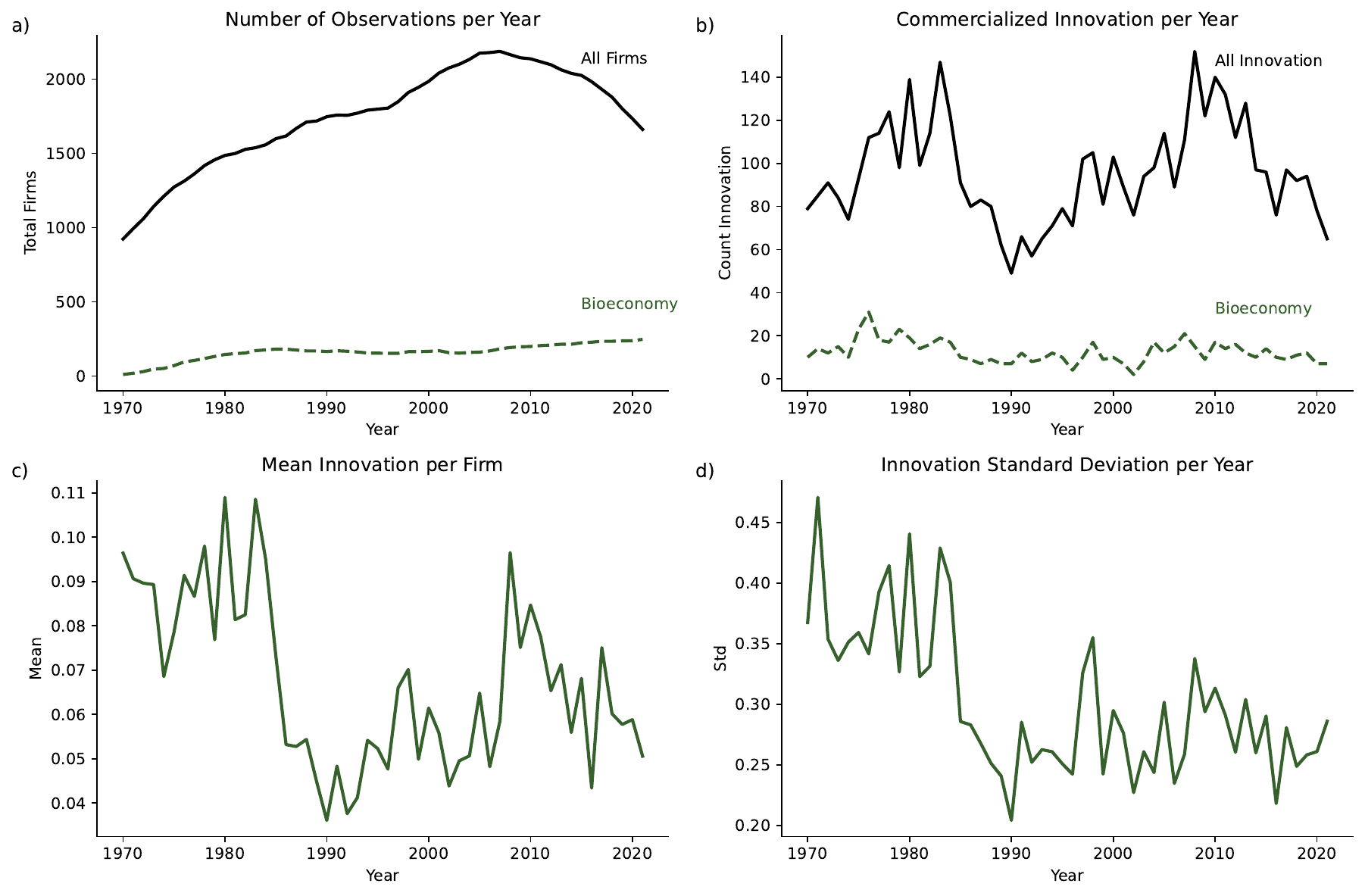}}

}

\caption{\label{fig-panel-composition}\textbf{Panel Composition and
Innovation Output Over Time.} Number of active firms (panel a), total
annual innovations (panel b), mean innovation rate per firm (panel c),
and standard deviation (panel d). Bioeconomy firms (dashed) and total
firms (solid).}

\end{figure}

Figure~\ref{fig-panel-composition} shows panel structure and innovation
dynamics over time. The number of active firms (panel a) grows steadily
between 1970, peaks around 2,000 firms in the 2000s, then declines
again. In total, we observed 3,739 unique firms. The number of active
bioeconomy firms remains more stable over the full period, totalling
578. Total innovations commercialized annually (panel b) fluctuate
between 60-140, with no clear secular trend, though bioeconomy
innovations remain consistently low (averaging 12.56 (SD=5.30)
innovations per year). Mean innovation rates per firm (panel c) decline
after 1980 from approximately 0.10 to 0.05-0.07 in later decades. This
decline is partly mechanical due to the expanding denominator of active
firms. The standard deviation (panel d) shows similar patterns,
indicating that the decline in mean rates reflects broad changes rather
than compositional shifts among top innovators.

\subsubsection{Model Specification}\label{model-specification}

We modeled the number of innovations commercialized by an organization
\(i\) in year \(t\) as following a Poisson distribution:

\[\log(\lambda_{it}) = \sum_{k=1}^{5} \beta_k X^k_{it-1} + \sum_{k=1}^{5} \theta_k (X^k_{it-1} \times {\text{Bioeconomy}_{it-1}}) + \gamma \text{Age}_{it-1} + \delta_t + \varepsilon_{it}\]

where \(\lambda_{it}\) is the predicted count of innovations, \(k\)
indexes the five network and cognitive proximity variables: two-step
betweenness, direct ties, indirect ties, cognitive proximity, and
(cognitive proximity)\(^2\), and \(\text{Bioeconomy}_{it-1}\) is an
indicator equal to 1 if organization \(i\)'s cumulative innovation
output in \(t-1\) was at least 25\% bioeconomy-related.

\(\text{Age}_{it-1}\) is the firm age control, and \(\delta_t\)
represents year fixed effects to account for common time-varying factors
affecting all firms. All explanatory variables except year fixed effects
are lagged one period to mitigate endogeneity concerns. Variables thus
measure network position and cognitive proximity at the beginning of the
year in which innovations are commercialized.

In addition to this main model, we considered ties, brokerage, and
cognitive proximity individually, as well as excluding year fixed
effects.

\section{Results}\label{sec-results}

We present the results in two parts. First, we provide brief qualitative
context of the network structure, the position of selected bioeconomy
actors within, and examples of their collaborations. Then, we turn to
the panel regression results. We report coefficients in
Table~\ref{tbl-panel-main-results} and interpret economic magnitudes
with the use of predicted innovation outputs in
Figure~\ref{fig-predictions}.

\subsection{Network Structure and Collaboration
Patterns}\label{sec-results-qual}

\begin{figure}

\centering{

\pandocbounded{\includegraphics[keepaspectratio]{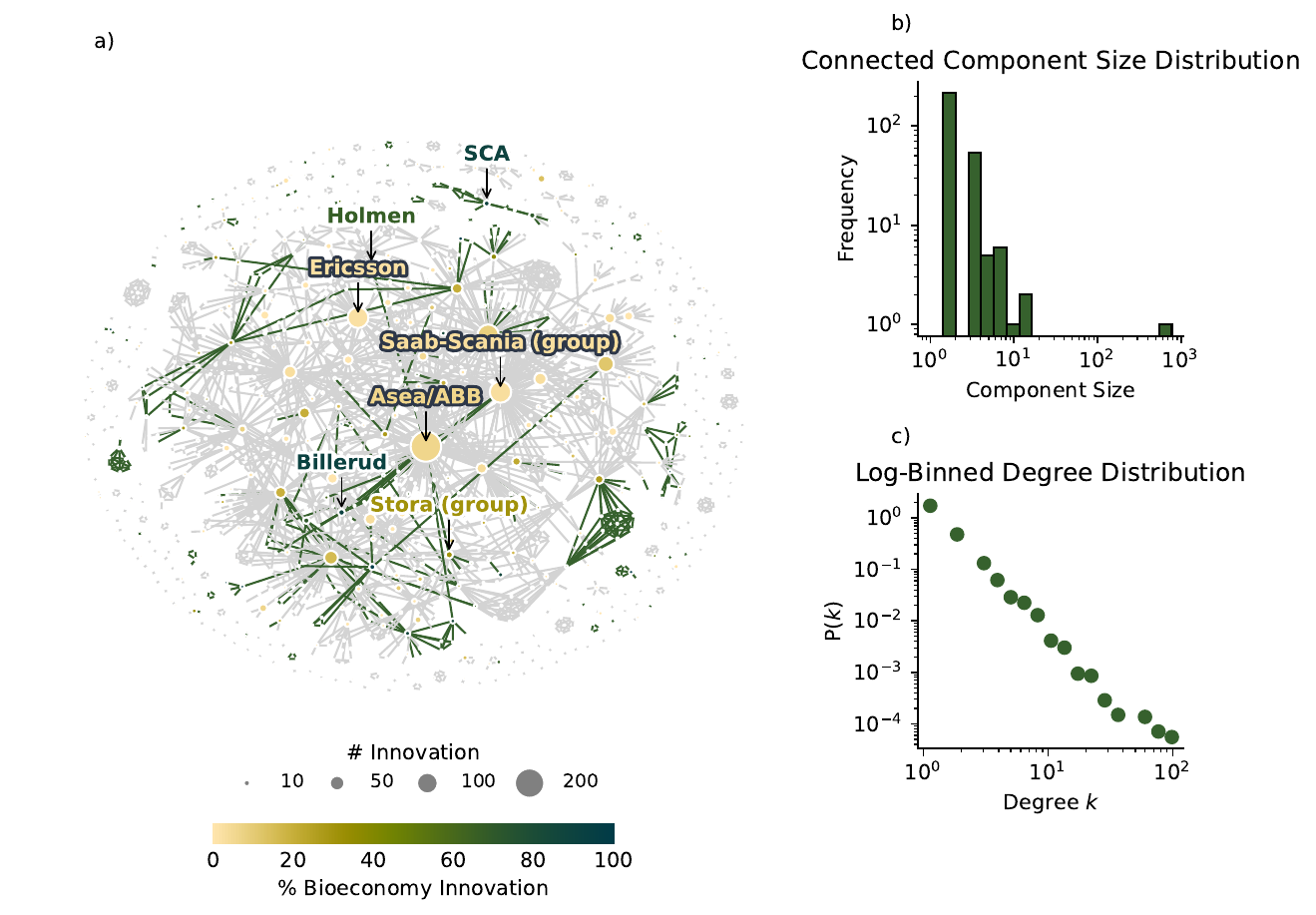}}

}

\caption{\label{fig-network}\textbf{Sweden's Innovation Collaboration
Network (1970--2021)}. Panel a) depicts all collaborations observed in
our study period; panel b) depicts component size distributions and
panel c) the log-binned degree distribution. Isolated nodes are omitted.
Green edges represent bioeconomy collaborations, gray edges other
collaborations.}

\end{figure}

Sweden's forest bioeconomy collaboration network exhibits sparse
connectivity and high fragmentation. It included 244 nodes which engaged
in at least one collaborative innovation output, with an average degree
of 2.41 (SD=2.08) (Table~\ref{tbl-network-descriptives}). Despite
bioeconomy actors commercializing fewer innovations, they had more
direct ties than actors who did not meet the bioeconomy-firm threshold
(Table~\ref{tbl-panel-desc}). Additionally, actors who collaborated
within the bioeconomy at least once represent important bridges in
Sweden's overall collaboration network
(Table~\ref{tbl-network-descriptives} and Figure~\ref{fig-network}
a).\footnote{An interactive version of the network depicted in
  Figure~\ref{fig-network} a) is available at
  \href{(https://pjkreutzer.github.io/files/bioeconomy_network)}{pjkreutzer.github.io/files/bioeconomy\_network}.}

Yet, major forestry organizations, such as SCA, Holmen, Stora and
Billerud, occupy rather peripheral positions in the network
(Figure~\ref{fig-network} a). Notably, these actors avoided direct
collaboration with one another. Thus, the role of network builders in
the Swedish bioeconomy falls on large engineering firms (ABB,
SAAB-Scania), with only low bioeconomy innovation intensity, and
universities such as Chalmers and KTH.

Analyzing the innovation descriptions included in the database revealed
that collaborations of the largest bioeconomy actors concentrated
heavily on systems and process-measurement for biomass flow, integrated
technological systems for machinery, and production systems for biomass
conversion and storage. Novel products capable of replacing fossil-based
materials were rare collaboration outcomes, despite industry rhetoric
emphasizing such transformative innovations as the explicit goal of
increased collaboration.

\subsection{Regression Results}\label{sec-results-panel}

Overall, our regression results were very robust across different
specifications (Table~\ref{tbl-panel-main-results} --
Table~\ref{tbl-panel-network-robustness}). Here, we focus on the main
specification in Table~\ref{tbl-panel-main-results}, which included all
covariates and year fixed effects (column 8).

\subsubsection{Direct Ties Were Associated with Increased Innovation
Output}\label{direct-ties-were-associated-with-increased-innovation-output}

The relationship between direct ties and innovation output showed clear
positive patterns, supporting \(H1a\). Incorporating the full
information of the panel regression indicated no systematic difference
in this relationship for bioeconomy and non-bioeconomy firms: having
more direct collaboration partners increases the amount of subsequent
innovation for both. However, Figure~\ref{fig-predictions} a) shows that
the predicted number of commercialized innovation per year for
non-bioeconomy firms moves from 0.067 at zero direct ties to 0.077 at
one direct tie, to 0.150 at the 99th percentile value of 6 direct ties.
Notably, bioeconomy firms exhibit the same pattern, albeit at a lower
overall level, entering the collaboration network increases the
predicted output from 0.039 to 0.048 at one direct tie, to 0.120 at the
99th percentile value of 6 direct ties.

These increases, especially at lower levels, may appear modest, but
translate to substantially more innovation output. On average, a
one-unit increase in direct ties increases the expected innovation
output per firm and per year by 14\% on average. However, the
uncertainty of the predicted innovation output increases quickly, owing
to the fact that few observations had more than a few direct ties. Thus,
our results suggest that promoting collaboration for otherwise isolated
notes may have more certain returns than increasing collaboration ties
for already connected actors.

\subsubsection{No Clear Evidence that Indirect Ties Affect Innovation
Output}\label{no-clear-evidence-that-indirect-ties-affect-innovation-output}

Our results provide no support for \emph{H1b}. We cannot reject the null
hypothesis of no effect for indirect ties on innovation output.
Figure~\ref{fig-predictions} b) suggests that there may be an
association different from zero for small values of indirect
connections, but as was the case for direct ties, as the number of
indirect ties increases, so does the uncertainty of the effect---in this
case even until it becomes impossible to distinguish it form zero.
Additionally, the association remains essentially flat for both
bioeconomy and non-bioeconomy firms. This suggests that collaboration
partners' connections contribute little, if anything, to innovation
productivity. Direct collaboration, not passive position appears to be
the more important driver of innovation output.

\subsubsection{Brokerage Positions Showed No Clear Association with
Innovation
Output}\label{brokerage-positions-showed-no-clear-association-with-innovation-output}

Two-step betweenness---measuring how often a node serves as the
exclusive bridge between otherwise disconnected collaborators---showed
no clear relationship with innovation output, contradicting hypothesis
\emph{H1c}. The predicted innovation counts in
Figure~\ref{fig-predictions} c) are nearly flat and confidence intervals
wide, indicating substantial uncertainty about any effect. This means
that we cannot make a precise statement regarding the relationship
between connecting otherwise disconnected groups and innovation output.
Our results do suggest, however, that this relationship did not differ
substantially for actors focused on bioeconomy innovation.

\subsubsection{Cognitive Proximity Showed Predicted Inverse-U, but with
Modest
Effects}\label{cognitive-proximity-showed-predicted-inverse-u-but-with-modest-effects}

Our evidence regarding cognitive proximity is mixed. We find the
predicted inverse-U relationship (supporting \emph{H1d}), but reject
both \emph{H2a} and \emph{H2b}: bioeconomy firms show higher average
proximity (0.09 vs 0.05) but remain below, not above, optimal levels.

\begin{figure}

\centering{

\pandocbounded{\includegraphics[keepaspectratio]{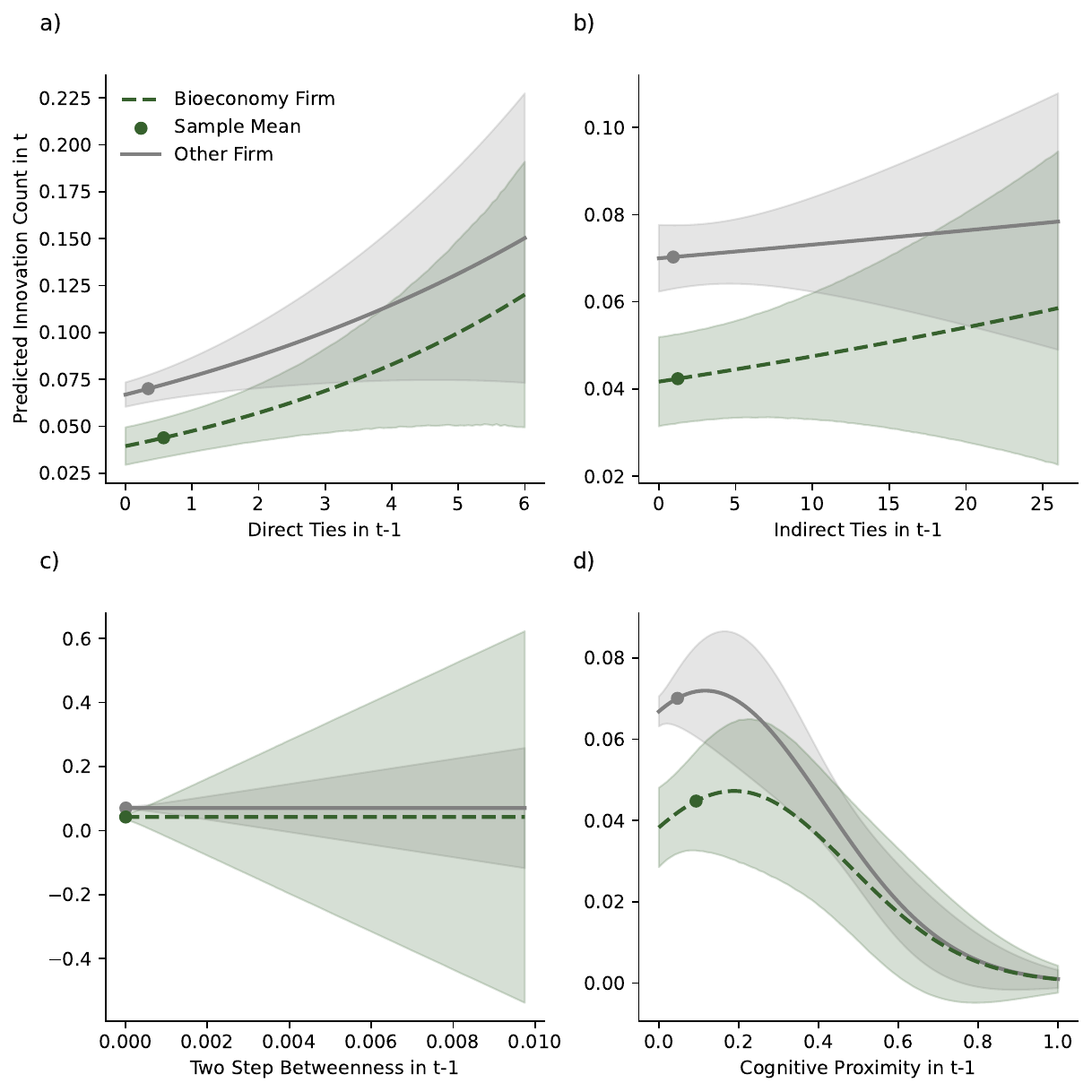}}

}

\caption{\label{fig-predictions}\textbf{Predictions of Subsequent
Innovation by Network and Knowledge Predictors.} Predicted innovation
counts comparing bioeconomy firms (dashed green) and non-bioeconomy
firms (solid gray). Panel a) direct ties (nodes distance 1), panel b)
indirect ties (nodes distance 2), panel c) two-step betweenness, panel
d) cognitive proximity. Dots show sample means, shaded areas show 95\%
confidence intervals. Based on Poisson model including year FE and
standard errors clustered by firm (column 8 in
Table~\ref{tbl-panel-main-results}). Panel a) and b) cover the 99th
precentile of sample values, to avoid distortion from outliers; panel c)
displays the full empirically observed range; and panel d) the
theoretically possible range of the cognitive proximity variable. Other
covariates held at representative values.}

\end{figure}

Figure~\ref{fig-predictions} d) shows the inverted-U clearly, with a
maximum around Jaccard-index 0.20. However, effect sizes are modest.
Moving from observed mean proximity (0.05) to the optimum (0.20)
increases predicted innovations by only 0.001 per year for
non-bioeconomy firms. For bioeconomy firms, moving from their observed
mean (0.09) to their optimum increases output by approximately 0.002 per
year. Moreover, few firms achieve proximity levels high enough for
negative returns to emerge.

These results contradict the hypothesis that bioeconomy actors suffer
from excessive cognitive lock-in. Instead, slightly greater similarity
with collaboration partners could marginally benefit innovation, though
practical gains appear limited.

\subsubsection{Bioeconomy Firms Followed Similar Mechanics but Were
Overall Less
Innovative}\label{bioeconomy-firms-followed-similar-mechanics-but-were-overall-less-innovative}

Across all measures, bioeconomy-intense firms showed consistently lower
innovation output than non-bioeconomy firms. However, associations
between network positions and cognitive proximity were not different
from those of non-bioeconomy firms. This suggests that bioeconomy firms
operate with similar collaboration mechanics as other firms. Output
deficits cannot meaningfully be credited to differences we observed in
this study, but may instead lie in unobserved bioeconomy-specific
barriers to innovation.

Previous literature has identified tradition-focused, risk-averse firm
dispositions in forestry industries as one candidate for these barriers
(\citeproc{ref-lamberg2017InstitutionalPathDependence}{Lamberg et al.,
2017}). Others have argued that core bioeconomy sectors (specifically
pulp and paper) are mature, concentrated and technologically saturated
(\citeproc{ref-ojala2012EvolutionGlobalPaper}{Ojala et al., 2012}). But
recent work also finds evidence that these sectors have entered a new
life cycle phase in Sweden, in which incumbents extend existing
knowledge into new product categories
(\citeproc{ref-kreutzer2025BioeconomyNewLife}{Kreutzer, 2025}).
Identifying the strength and interaction of these dynamics presents an
important avenue for future research.

\begin{table}

\caption{\label{tbl-panel-main-results}Main Panel Results from Poisson}

\centering{

\begin{adjustbox}{max width=\textwidth}
\begin{tabular}{@{\extracolsep{5pt}}lcccccccc}
\toprule
& \multicolumn{8}{c}{\textit{Dependent variable: Subsequent Innovation Count}} \
\cr \cmidrule(lr){2-9}
 & \multicolumn{8}{c}{} \\ & (1) & (2) & (3) & (4) & (5) & (6) & (7) & (8) \\
\midrule
 Bioeconomy Firm & -0.592$^{***}$ & -0.567$^{***}$ & -0.497$^{***}$ & -0.478$^{***}$ & -0.639$^{***}$ & -0.611$^{***}$ & -0.603$^{***}$ & -0.584$^{***}$ \\
& (0.119) & (0.118) & (0.107) & (0.107) & (0.139) & (0.138) & (0.133) & (0.132) \\
 Direct Ties & 0.133$^{***}$ & 0.134$^{***}$ & & & & & 0.135$^{***}$ & 0.135$^{***}$ \\
& (0.016) & (0.015) & & & & & (0.034) & (0.042) \\
 Indirect Ties & 0.004$^{}$ & 0.006$^{}$ & & & & & 0.005$^{}$ & 0.004$^{}$ \\
& (0.006) & (0.006) & & & & & (0.007) & (0.008) \\
 2-step Betweenness & & & 530.363$^{***}$ & 533.259$^{***}$ & & & -0.148$^{}$ & -0.221$^{}$ \\
& & & (24.396) & (18.566) & & & (110.118) & (142.683) \\
 Cognitive proximity & & & & & 6.836$^{***}$ & 7.416$^{***}$ & 0.621$^{}$ & 1.273$^{}$ \\
& & & & & (2.255) & (2.378) & (0.938) & (0.977) \\
$\text{Cognitive Proximity}^{2}$ & & & & & -16.118$^{***}$ & -16.915$^{***}$ & -4.675$^{**}$ & -5.481$^{***}$ \\
& & & & & (5.548) & (5.775) & (1.919) & (2.047) \\
 Direct Ties $\times$ Bioeconomy Firm & 0.026$^{}$ & 0.045$^{}$ & & & & & 0.038$^{}$ & 0.051$^{}$ \\
& (0.038) & (0.038) & & & & & (0.068) & (0.067) \\
 Indirect Ties $\times$ Bioeconomy Firm & 0.017$^{}$ & 0.015$^{}$ & & & & & 0.008$^{}$ & 0.009$^{}$ \\
& (0.013) & (0.013) & & & & & (0.014) & (0.015) \\
 2-step Betweenness $\times$ Bioeconomy Firm & & & 4.006$^{}$ & 4.011$^{}$ & & & 0.001$^{}$ & 0.001$^{}$ \\
& & & (6812.862) & (6250.848) & & & (1054.659) & (721.338) \\
 Cognitive proximity $\times$ Bioeconomy Firm & & & & & 1.250$^{}$ & 0.811$^{}$ & 1.106$^{}$ & 0.964$^{}$ \\
& & & & & (3.489) & (3.524) & (1.713) & (1.833) \\
$\text{Cognitive Proximity}^{2}$ $\times$ Bioeconomy Firm & & & & & -1.092$^{}$ & -0.292$^{}$ & -0.387$^{}$ & -0.449$^{}$ \\
& & & & & (9.562) & (9.534) & (3.222) & (3.663) \\
 Intercept & -2.924$^{***}$ & -2.510$^{***}$ & -2.863$^{***}$ & -2.533$^{***}$ & -2.867$^{***}$ & -2.475$^{***}$ & -2.869$^{***}$ & -2.558$^{***}$ \\
& (0.019) & (0.155) & (0.024) & (0.152) & (0.020) & (0.146) & (0.018) & (0.165) \\
\addlinespace
 Time Effects & No & Yes & No & Yes & No & Yes & No & Yes \\
 Age Control & Yes & Yes & Yes & Yes & Yes & Yes & Yes & Yes \\
\midrule
 N Firms & 3,739 & 3,739 & 3,739 & 3,739 & 3,739 & 3,739 & 3,739 & 3,739 \\
 Observations & 89931 & 89931 & 89931 & 89931 & 89931 & 89931 & 89931 & 89931 \\
 Pseudo $R^2$ & 0.049 & 0.060 & 0.034 & 0.044 & 0.024 & 0.034 & 0.055 & 0.065 \\
\midrule
\textit{Note:} & \multicolumn{8}{r}{$^{*}$p$<$0.1; $^{**}$p$<$0.05; $^{***}$p$<$0.01} \\
\multicolumn{9}{r}\textit{Standard errors clustered at the firm level in parentheses.} \\
\bottomrule
\end{tabular}
\end{adjustbox}

}

\end{table}%

\subsection{Endogenity Sensitivity
Checks}\label{endogenity-sensitivity-checks}

One limitation of the above analysis is that, although specified with
time lags, it does not completely rule out the possibility that firm's
propensity to collaborate successfully may reflect latent innovation
capacity. To deal with potential endogeneity, we follow earlier
literature that used neighbors' characteristics as an instrument for
direct ties
(\citeproc{ref-bramoulle2009IdentificationPeerEffects}{Bramoullé et al.,
2009}; \citeproc{ref-zenou2025PeerVsNetwork}{Zenou, 2025}). The
underlying rationale is that neighbors' characteristics affect the
propensity of a focal firm to collaborate with them, while it does not
affect innovation output of the focal firm. Specifically, we use the
number of neighbors' neighbors, under the rationale that firms are more
likely to collaborate with well-connected firms
(\citeproc{ref-newman2001StructureScientificCollaboration}{Newman,
2001}).~

The results are~presented in Table~\ref{tbl-IV}. We run both linear
models and Poisson GMM models. The instrument pass the Kleibergen-Paap
tests and Cragg--Donald F tests for underidentification,~and Anderson LM
tests for weak instrument. The results show that model coefficients and
results for the baseline variables are very similar.

We include these results not as definitive proof against endogeneity
concerns, but rather as sensitivity checks of our main results. The
results suggest that the positive relationship between direct ties and
innovation output discussed above is fairly robust.

\begin{table}

\caption{\label{tbl-IV}Instrumental Variable Regression Endogeneity
Check}

\centering{

\begin{adjustbox}{max width=\textwidth}
\footnotesize
\begin{tabular}{l*{4}{D{.}{.}{-1}}}
\hline
                    &\multicolumn{1}{c}{(1)}&\multicolumn{1}{c}{(2)}&\multicolumn{1}{c}{(3)}&\multicolumn{1}{c}{(4)}\\
                    &\multicolumn{1}{c}{IV (Linear)}&\multicolumn{1}{c}{IV (Linear)}&\multicolumn{1}{c}{IV Poisson}&\multicolumn{1}{c}{IV Poisson}\\
\hline
      &                     &                     &                     &                     \\
Direct ties         &      0.0307\sym{***}&      0.0200\sym{***}&       0.154\sym{***}&       0.193\sym{***}\\
                    &   (0.00835)         &   (0.00569)         &   (0.00861)         &    (0.0195)         \\
\addlinespace
2-step Betweenness  &                     &       421.4\sym{***}&                     &      -241.1\sym{***}\\
                    &                     &     (113.5)         &                     &     (76.43)         \\
\addlinespace
Cognitive Proximity &                     &      -0.151\sym{***}&                     &       0.479         \\
                    &                     &    (0.0478)         &                     &     (1.007)         \\
\addlinespace
Cognitive Proximity$^2$&                     &      0.0625         &                     &      -4.373\sym{**} \\
                    &                     &    (0.0444)         &                     &     (1.865)         \\
\addlinespace
Constant            &      0.0796\sym{***}&      0.0813\sym{***}&      -2.511\sym{***}&      -2.510\sym{***}\\
                    &    (0.0135)         &    (0.0138)         &     (0.156)         &     (0.156)         \\
\addlinespace
Time Effects        &         \text{Yes}         &         \text{Yes}         &         \text{Yes}         &         \text{Yes}         \\
Age Control        &         \text{Yes}         &         \text{Yes}         &         \text{Yes}         &         \text{Yes}         \\
\hline
N Firms        &         3,739         &         3,739         &         3,739         &         3,739         \\
Observations        &         89,931         &         89,931         &         89,931         &         89,931         \\
\hline
\multicolumn{5}{l}{\footnotesize Standard errors in parentheses}\\
\multicolumn{5}{l}{\footnotesize \sym{*} \(p<0.10\), \sym{**} \(p<0.05\), \sym{***} \(p<0.01\)}\\
\end{tabular}
\end{adjustbox}

}

\end{table}%

\section{Conclusion}\label{sec-conclusion}

We analyzed fifty years of collaboration behavior and associated
innovation output for a new panel of actors commercializing innovation
in Sweden between 1970 and 2021. Thus we contribute a comprehensive
examination of collaboration and actual innovation output, capturing the
emerging and cross-sectoral bioeconomy within the Swedish innovation
system.

We find robust evidence that direct collaboration ties are positively
associated with subsequent innovation output. This relationship has
substantive meaning. Entering the collaboration network was associated
with a 0.010 (somewhat less for bioeconomy actors at 0.008) increase in
a firm's annual innovation output. At the individual firm level this may
appear small, however, across the many isolated innovators in Sweden's
innovation system, increasing the collaboration network's density would
translate into substantially more innovation per year.

We find no evidence that supports positive relationships between
increasing indirect connections, or connecting otherwise separate parts
of the network and innovation production, once we control for direct
ties.

Cognitive proximity follows the theorized inverted-U relationship with
innovation, but reaching optimal cognitive proximity to network partners
was associated with negligible increases in innovation production.
Moreover, we find no evidence that substantiates concerns about
excessive similarity among bioeconomy collaborators in the Swedish
context.

Critically, bioeconomy actors are not subject to different mechanics in
the relationship of collaboration and innovation performance. Bioeconomy
actors are not failing to benefit from collaboration, they appear simply
less innovative overall. The bioeconomy involves firms which take on
central network building roles in Sweden's overall collaboration
network, but the biggest actors of core sectors avoid collaborating
directly with one another.

\subsection{Policy Implications}\label{policy-implications}

These findings carry important implications for policymakers seeking to
strengthen the Swedish bioeconomy, and accelerating its transition.

First, policies should prioritize increasing the quantity of
collaborations rather than optimizing their composition. Current focus
in policy discourse on identifying the right partners or fostering
specific cross-sectoral linkages may be misplaced when it comes to
increasing innovation output. Rather, lowering barriers to collaboration
may yield greater and more certain returns.

Second, such policies need not be exclusively tailored to the
bioeconomy. Bioeconomy and non-bioeconomy firms appear to follow the
same mechanism between collaboration and innovation output. Promoting
collaboration across the Swedish innovation system, while less specific,
may be as effective for the bioeconomy and simultaneously increase
innovation output in other sectors. Considering the large number of
isolated bioeconomy and non-bioeconomy firms in the network, such
efforts could instead target first-time collaborators.

\subsection{Limitations and Future Research}\label{sec-limitations}

Several limitations qualify these conclusions. Our classification of
bioeconomy actors relies on a threshold criterion (25\% bioeconomy
innovation share) which, while pragmatic, lacks strong theoretical
foundation. Robustness checks, however, do not suggest that our results
are sensitive to this choice. Further, we only consider collaborations
where independent actors work together resulting in a commercialized
innovation, at the expense of other meaningful forms of knowledge and
resource exchange, such as joint ventures, industry and research
consortia, or informal networks
(\citeproc{ref-foschi2025UnderstandingInterOrganizationalDynamics}{Foschi
et al., 2025};
\citeproc{ref-soderholm2012FirmCollaborationEnvironmental}{Söderholm and
Bergquist, 2012}). Our data only captures collaborations that succeeded
in bringing an innovation to market. Failed attempts may nevertheless
generate learning.

Finally, we focused on aggregated innovation output. While such counts
provide meaningful insights into the development of technological
change, they largely omit technological diffusion and outcomes. Few
radical innovations could provide important stimuli to the bioeconomy
and, indeed economy at large.

Future research should investigate these dynamics. Additionally, it
would be fruitful to identify reasons behind bioeconomy incumbents
peripheral position despite apparent benefits from collaboration.
Further, other barriers to innovation in the bioeconomy than lacking
collaboration should receive more attention. Advances in these
directions could come from additional quantitative or qualitative work;
comparative perspectives could further help illuminate the relative
importance of factors in different economic and institutional settings.

\subsection{Concluding Remarks}\label{concluding-remarks}

The bioeconomy transition requires integrating traditional biomass
sectors with diverse downstream industries---a challenge that
collaboration is widely expected to address. Our findings both validate
and qualify this expectation. Collaboration does indeed enhance
innovation output, and cognitive proximity indeed has a sweet spot.
However, the solution appears less about finding optimal partners than
about collaborating more broadly. For policymakers, the implication is
clear: focus on stimulating collaboration quantity and reducing barriers
to partnership formation, rather than attempting to engineer specific
network configurations. In an innovation system where most actors remain
isolated, the marginal value of any new collaboration is likely to
exceed the marginal value of optimizing existing ones.

\section*{Code Availability}\label{sec-code-availability}
\addcontentsline{toc}{section}{Code Availability}

The analysis of the network was carried out using the package
\texttt{networkx}
(\citeproc{ref-hagberg2008ExploringNetworkStructure}{Hagberg et al.,
2008}) and the polars package
(\citeproc{ref-vink2024PolarsPolarsPython}{Vink et al., 2024}) in
Python. Statistical modelling was conducted using the
\texttt{statsmodels} (\citeproc{ref-seabold2010statsmodels}{Seabold and
Perktold, 2010}) and \texttt{marginaleffects}
(\citeproc{ref-arel-bundock2024HowInterpretStatistical}{Arel-Bundock et
al., 2024}) package. All code is available at
\href{github.com/pjkreutzer/swinno_bioeconomy_network}{\texttt{github.com/pjkreutzer/swinno\_bioeconomy\_network}}.

\section*{References}\label{references}
\addcontentsline{toc}{section}{References}

\phantomsection\label{refs}

\section{Appendix}\label{appendix}

\begin{longtable}[]{@{}
  >{\raggedright\arraybackslash}p{(\linewidth - 2\tabcolsep) * \real{0.5000}}
  >{\raggedright\arraybackslash}p{(\linewidth - 2\tabcolsep) * \real{0.5000}}@{}}
\caption{Sectors and keywords used in bioeconomy innovation
queries}\label{tbl-query}\tabularnewline
\toprule\noalign{}
\begin{minipage}[b]{\linewidth}\raggedright
SNI Code -- Sector
\end{minipage} & \begin{minipage}[b]{\linewidth}\raggedright
Keywords used in Swedish
\end{minipage} \\
\midrule\noalign{}
\endfirsthead
\toprule\noalign{}
\begin{minipage}[b]{\linewidth}\raggedright
SNI Code -- Sector
\end{minipage} & \begin{minipage}[b]{\linewidth}\raggedright
Keywords used in Swedish
\end{minipage} \\
\midrule\noalign{}
\endhead
\bottomrule\noalign{}
\endlastfoot
02 -- Forestry and related services &
\multirow{6}{=}{\begin{minipage}[t]{\linewidth}\raggedright
timber (virke); cellulose (cellulos); lignin (lignin); chip (spån); bark
(bark); levulinic acid (levulinsyra); furfural (furfural); black tar
(svarttjära); black liquor (svartlut); plant-based (växtbas); wood
(ved); timber (trä); forest (skog); biofuel (biobränsle); biological
(biologiskt); biodegradable (nedbrytbar); paper (papper); cardboard
(pappret); carton (karton); lyocell (lyocell); Tencel (tencel)
\end{minipage}} \\
20 -- Wood and wood product manufacturing except furniture \\
21 -- Pulp, paper and paper product manufacturing \\
36 -- Furniture manufacturing; other manufacturing \\
 \\
 \\
& \\
& \\
\end{longtable}

\subsection{Robustness Checks}\label{robustness-checks}

\begin{table}

\caption{\label{tbl-firm-robustness}Collaborators Only Panel Results
from Poisson}

\centering{

\begin{adjustbox}{max width=\textwidth}
\begin{tabular}{@{\extracolsep{5pt}}lcccccccc}
\toprule
& \multicolumn{8}{c}{\textit{Dependent variable: Subsequent Innovation Count}} \
\cr \cmidrule(lr){2-9}
 & \multicolumn{8}{c}{} \\ & (1) & (2) & (3) & (4) & (5) & (6) & (7) & (8) \\
\midrule
 Bioeconomy Firm & -0.173$^{}$ & -0.143$^{}$ & -0.158$^{}$ & -0.123$^{}$ & -0.037$^{}$ & 0.004$^{}$ & -0.035$^{}$ & -0.028$^{}$ \\
& (0.161) & (0.154) & (0.140) & (0.137) & (0.193) & (0.182) & (0.190) & (0.180) \\
 Direct Ties & 0.131$^{***}$ & 0.132$^{***}$ & & & & & 0.134$^{***}$ & 0.133$^{***}$ \\
& (0.015) & (0.015) & & & & & (0.032) & (0.042) \\
 Indirect Ties & 0.001$^{}$ & 0.004$^{}$ & & & & & 0.003$^{}$ & 0.004$^{}$ \\
& (0.006) & (0.006) & & & & & (0.007) & (0.008) \\
 2-step Betweenness & & & 497.334$^{***}$ & 513.090$^{***}$ & & & -1.273$^{}$ & -0.589$^{}$ \\
& & & (26.674) & (20.479) & & & (103.277) & (144.204) \\
 Cognitive proximity & & & & & 5.219$^{**}$ & 6.087$^{***}$ & -0.782$^{}$ & -0.076$^{}$ \\
& & & & & (2.059) & (2.291) & (0.835) & (0.864) \\
$\text{Cognitive Proximity}^{2}$ & & & & & -13.953$^{***}$ & -15.238$^{***}$ & -2.874$^{*}$ & -3.565$^{**}$ \\
& & & & & (5.155) & (5.584) & (1.594) & (1.687) \\
 Direct Ties $\times$ Bioeconomy Firm & -0.046$^{}$ & -0.019$^{}$ & & & & & 0.014$^{}$ & 0.028$^{}$ \\
& (0.045) & (0.043) & & & & & (0.065) & (0.064) \\
 Indirect Ties $\times$ Bioeconomy Firm & 0.016$^{}$ & 0.014$^{}$ & & & & & 0.003$^{}$ & 0.003$^{}$ \\
& (0.012) & (0.012) & & & & & (0.014) & (0.014) \\
 2-step Betweenness $\times$ Bioeconomy Firm & & & 3.431$^{}$ & 3.505$^{}$ & & & 0.015$^{}$ & 0.004$^{}$ \\
& & & (3636.394) & (3346.048) & & & (973.365) & (621.702) \\
 Cognitive proximity $\times$ Bioeconomy Firm & & & & & -4.512$^{*}$ & -4.985$^{*}$ & -1.219$^{}$ & -0.747$^{}$ \\
& & & & & (2.566) & (2.682) & (1.610) & (1.770) \\
$\text{Cognitive Proximity}^{2}$ $\times$ Bioeconomy Firm & & & & & 9.247$^{}$ & 10.003$^{}$ & 1.730$^{}$ & 0.466$^{}$ \\
& & & & & (5.938) & (6.289) & (2.593) & (3.271) \\
 Intercept & -2.852$^{***}$ & -2.383$^{***}$ & -2.716$^{***}$ & -2.407$^{***}$ & -2.658$^{***}$ & -2.395$^{***}$ & -2.681$^{***}$ & -2.331$^{***}$ \\
& (0.040) & (0.250) & (0.053) & (0.245) & (0.043) & (0.248) & (0.037) & (0.240) \\
\addlinespace
 Time Effects & No & Yes & No & Yes & No & Yes & No & Yes \\
 Age Control & Yes & Yes & Yes & Yes & Yes & Yes & Yes & Yes \\
\midrule
 N Firms & 1,464 & 1,464 & 1,464 & 1,464 & 1,464 & 1,464 & 1,464 & 1,464 \\
 Observations & 38834 & 38834 & 38834 & 38834 & 38834 & 38834 & 38834 & 38834 \\
 Pseudo $R^2$ & 0.079 & 0.099 & 0.061 & 0.077 & 0.043 & 0.058 & 0.096 & 0.112 \\
\midrule
\textit{Note:} & \multicolumn{8}{r}{$^{*}$p$<$0.1; $^{**}$p$<$0.05; $^{***}$p$<$0.01} \\
\multicolumn{9}{r}\textit{Standard errors clustered at the firm level in parentheses.} \\
\bottomrule
\end{tabular}
\end{adjustbox}

}

\end{table}%

\begin{table}

\caption{\label{tbl-panel-threshold-robustness}All Firms Panel Results
from Poisson---Bioeconomy Threshold = 50\%}

\centering{

\begin{adjustbox}{max width=\textwidth}
\begin{tabular}{@{\extracolsep{5pt}}lcccccccc}
\toprule
& \multicolumn{8}{c}{\textit{Dependent variable: Subsequent Innovation Count}} \
\cr \cmidrule(lr){2-9}
 & \multicolumn{8}{c}{} \\ & (1) & (2) & (3) & (4) & (5) & (6) & (7) & (8) \\
\midrule
 Bioeconomy Firm & -0.864$^{***}$ & -0.842$^{***}$ & -0.789$^{***}$ & -0.769$^{***}$ & -0.941$^{***}$ & -0.904$^{***}$ & -0.868$^{***}$ & -0.846$^{***}$ \\
& (0.110) & (0.108) & (0.106) & (0.105) & (0.118) & (0.115) & (0.109) & (0.108) \\
 Direct Ties & 0.132$^{***}$ & 0.133$^{***}$ & & & & & 0.134$^{***}$ & 0.134$^{***}$ \\
& (0.015) & (0.015) & & & & & (0.033) & (0.041) \\
 Indirect Ties & 0.005$^{}$ & 0.007$^{}$ & & & & & 0.005$^{}$ & 0.005$^{}$ \\
& (0.006) & (0.006) & & & & & (0.007) & (0.008) \\
 2-step Betweenness & & & 529.893$^{***}$ & 532.744$^{***}$ & & & -0.168$^{}$ & -0.265$^{}$ \\
& & & (24.333) & (18.504) & & & (105.921) & (137.513) \\
 Cognitive proximity & & & & & 6.864$^{***}$ & 7.462$^{***}$ & 0.677$^{}$ & 1.311$^{}$ \\
& & & & & (2.165) & (2.288) & (0.912) & (0.951) \\
$\text{Cognitive Proximity}^{2}$ & & & & & -16.177$^{***}$ & -17.021$^{***}$ & -4.723$^{**}$ & -5.517$^{***}$ \\
& & & & & (5.346) & (5.580) & (1.865) & (1.994) \\
 Direct Ties $\times$ Bioeconomy Firm & 0.052$^{}$ & 0.079$^{}$ & & & & & 0.079$^{}$ & 0.099$^{}$ \\
& (0.051) & (0.050) & & & & & (0.071) & (0.074) \\
 Indirect Ties $\times$ Bioeconomy Firm & 0.015$^{}$ & 0.012$^{}$ & & & & & 0.004$^{}$ & 0.003$^{}$ \\
& (0.017) & (0.016) & & & & & (0.019) & (0.019) \\
 2-step Betweenness $\times$ Bioeconomy Firm & & & 0.833$^{}$ & 0.842$^{}$ & & & 0.001$^{}$ & 0.001$^{}$ \\
& & & (10904.536) & (11777.173) & & & (2027.278) & (2022.117) \\
 Cognitive proximity $\times$ Bioeconomy Firm & & & & & 1.455$^{}$ & 0.686$^{}$ & 0.894$^{}$ & 0.700$^{}$ \\
& & & & & (3.593) & (3.604) & (1.983) & (2.115) \\
$\text{Cognitive Proximity}^{2}$ $\times$ Bioeconomy Firm & & & & & -0.523$^{}$ & 0.972$^{}$ & -0.080$^{}$ & -0.131$^{}$ \\
& & & & & (9.800) & (9.645) & (3.581) & (4.131) \\
 Intercept & -2.914$^{***}$ & -2.506$^{***}$ & -2.851$^{***}$ & -2.529$^{***}$ & -2.857$^{***}$ & -2.445$^{***}$ & -2.860$^{***}$ & -2.533$^{***}$ \\
& (0.018) & (0.156) & (0.023) & (0.152) & (0.020) & (0.142) & (0.017) & (0.161) \\
\addlinespace
 Time Effects & No & Yes & No & Yes & No & Yes & No & Yes \\
 Age Control & Yes & Yes & Yes & Yes & Yes & Yes & Yes & Yes \\
\midrule
 N Firms & 3,739 & 3,739 & 3,739 & 3,739 & 3,739 & 3,739 & 3,739 & 3,739 \\
 Observations & 89931 & 89931 & 89931 & 89931 & 89931 & 89931 & 89931 & 89931 \\
 Pseudo $R^2$ & 0.050 & 0.062 & 0.036 & 0.046 & 0.026 & 0.036 & 0.057 & 0.067 \\
\midrule
\textit{Note:} & \multicolumn{8}{r}{$^{*}$p$<$0.1; $^{**}$p$<$0.05; $^{***}$p$<$0.01} \\
\multicolumn{9}{r}\textit{Standard errors clustered at the firm level in parentheses.} \\
\bottomrule
\end{tabular}
\end{adjustbox}

}

\end{table}%

\begin{table}

\caption{\label{tbl-panel-network-robustness}All Firms Panel Results
from Poisson---Network Window = 20 Years}

\centering{

\begin{adjustbox}{max width=\textwidth}
\begin{tabular}{@{\extracolsep{5pt}}lcccccccc}
\toprule
& \multicolumn{8}{c}{\textit{Dependent variable: Subsequent Innovation Count}} \
\cr \cmidrule(lr){2-9}
 & \multicolumn{8}{c}{} \\ & (1) & (2) & (3) & (4) & (5) & (6) & (7) & (8) \\
\midrule
 Bioeconomy Firm & -0.592$^{***}$ & -0.567$^{***}$ & -0.497$^{***}$ & -0.478$^{***}$ & -0.639$^{***}$ & -0.611$^{***}$ & -0.603$^{***}$ & -0.584$^{***}$ \\
& (0.119) & (0.118) & (0.107) & (0.107) & (0.139) & (0.138) & (0.133) & (0.132) \\
 Direct Ties & 0.133$^{***}$ & 0.134$^{***}$ & & & & & 0.135$^{***}$ & 0.135$^{***}$ \\
& (0.016) & (0.015) & & & & & (0.034) & (0.042) \\
 Indirect Ties & 0.004$^{}$ & 0.006$^{}$ & & & & & 0.005$^{}$ & 0.004$^{}$ \\
& (0.006) & (0.006) & & & & & (0.007) & (0.008) \\
 2-step Betweenness & & & 530.363$^{***}$ & 533.259$^{***}$ & & & -0.148$^{}$ & -0.221$^{}$ \\
& & & (24.396) & (18.566) & & & (110.051) & (142.636) \\
 Cognitive proximity & & & & & 6.840$^{***}$ & 7.420$^{***}$ & 0.625$^{}$ & 1.277$^{}$ \\
& & & & & (2.255) & (2.378) & (0.938) & (0.977) \\
$\text{Cognitive Proximity}^{2}$ & & & & & -16.124$^{***}$ & -16.921$^{***}$ & -4.681$^{**}$ & -5.487$^{***}$ \\
& & & & & (5.546) & (5.774) & (1.920) & (2.047) \\
 Direct Ties $\times$ Bioeconomy Firm & 0.026$^{}$ & 0.045$^{}$ & & & & & 0.038$^{}$ & 0.051$^{}$ \\
& (0.038) & (0.038) & & & & & (0.068) & (0.067) \\
 Indirect Ties $\times$ Bioeconomy Firm & 0.017$^{}$ & 0.015$^{}$ & & & & & 0.008$^{}$ & 0.009$^{}$ \\
& (0.013) & (0.013) & & & & & (0.014) & (0.015) \\
 2-step Betweenness $\times$ Bioeconomy Firm & & & 4.006$^{}$ & 4.011$^{}$ & & & 0.001$^{}$ & 0.001$^{}$ \\
& & & (6812.862) & (6250.848) & & & (1054.725) & (721.341) \\
 Cognitive proximity $\times$ Bioeconomy Firm & & & & & 1.249$^{}$ & 0.809$^{}$ & 1.105$^{}$ & 0.963$^{}$ \\
& & & & & (3.490) & (3.524) & (1.713) & (1.834) \\
$\text{Cognitive Proximity}^{2}$ $\times$ Bioeconomy Firm & & & & & -1.092$^{}$ & -0.292$^{}$ & -0.387$^{}$ & -0.449$^{}$ \\
& & & & & (9.564) & (9.536) & (3.225) & (3.666) \\
 Intercept & -2.924$^{***}$ & -2.510$^{***}$ & -2.863$^{***}$ & -2.533$^{***}$ & -2.867$^{***}$ & -2.475$^{***}$ & -2.869$^{***}$ & -2.558$^{***}$ \\
& (0.019) & (0.155) & (0.024) & (0.152) & (0.020) & (0.146) & (0.018) & (0.165) \\
\addlinespace
 Time Effects & No & Yes & No & Yes & No & Yes & No & Yes \\
 Age Control & Yes & Yes & Yes & Yes & Yes & Yes & Yes & Yes \\
\midrule
 N Firms & 3,739 & 3,739 & 3,739 & 3,739 & 3,739 & 3,739 & 3,739 & 3,739 \\
 Observations & 89931 & 89931 & 89931 & 89931 & 89931 & 89931 & 89931 & 89931 \\
 Pseudo $R^2$ & 0.049 & 0.060 & 0.034 & 0.044 & 0.024 & 0.034 & 0.055 & 0.065 \\
\midrule
\textit{Note:} & \multicolumn{8}{r}{$^{*}$p$<$0.1; $^{**}$p$<$0.05; $^{***}$p$<$0.01} \\
\multicolumn{9}{r}\textit{Standard errors clustered at the firm level in parentheses.} \\
\bottomrule
\end{tabular}
\end{adjustbox}

}

\end{table}%

\end{document}